\documentstyle[12pt]{article}
\title{\bf Various Properties of\\
Compact QED and Confining Strings}
\author{D.V.ANTONOV 
\thanks{Phone: 0049-30-2093 7974; Fax: 0049-30-2093 7631; 
E-mail address: antonov@pha2.physik.hu-berlin.de}~ \thanks{  
On leave of absence from the Institute of Theoretical and Experimental 
Physics, B.Cheremushkinskaya 25, 117 218, Moscow, Russia.}\\ 
{\it Institut f\"ur Physik, Humboldt-Universit\"at zu 
Berlin,}\\
{\it Invalidenstrasse 110, D-10115, Berlin, Germany}\\}
\date{}
\begin{document}
\maketitle
\vspace{1cm}
\centerline{\bf {Abstract}}

\vspace{3mm}

The effects bringing about by the finiteness of the photon mass due 
to the Debye screening in the monopole gas in three-dimensional 
compact QED are studied. In this respect, a representation 
of the partition function of this theory  
as an integral over monopole densities is derived. Dual formulation 
of the Wilson loop yields a new theory of confining strings, which in 
the low-energy limit almost coincides with the one corresponding to 
the case when the photon is considered to be massless, 
whereas in the high-energy limit these two theories are quite different 
from each other. 
The confining string mass operator in the low-energy limit 
is also found, and its dependence on the volume of observation is studied.  

\vspace{3mm}

\vspace{3mm}

\newpage

{\large\bf 1. Introduction}
\vspace{3mm}

Theory of magnetic monopoles in compact QED both on a lattice and in the 
continuum limit has a long history [1-3]. 
In Ref. [2], it has been demonstrated 
that at large distances the partition function of  
3D compact QED is nothing else but the one of a gas of 
monopoles with a Coulomb interaction, which in notations of Ref. 
[2] reads

$${\cal Z}=\sum\limits_{N=1}^{+\infty}\sum\limits_{q_a=\pm 1}^{}
\frac{\zeta^N}{N!}
\prod\limits_{i=1}^N 
\int d{\bf z}_i\exp\left[-\frac{\pi}{2e^2}\sum\limits_{a\ne b}^{}
\frac{q_a q_b}{\left|{\bf z}_a-{\bf z}_b\right|}
\right]. \eqno (1)$$

Due to Ref. [2],  
Eq. (1) could be equivalently rewritten as a partition function of a
3D sine-Gordon model of a scalar field, which effectively substitutes 
an infinite number of monopoles,    

$${\cal Z}=\int {\cal D}\chi\exp\left\{-\int d{\bf x}\left[\frac12 
(\partial_\mu\chi)^2-2\zeta\cos\left(\frac{2\pi}{e}
\chi\right)\right]\right\}. \eqno (2) $$

Universal Confining String Theory (UCST), proposed in Ref. [3], is a 
dual formulation of the Wilson loop in the model (2). This 
formulation enables one to rewrite the interaction of the field 
$\chi$ with a solid angle, formed by the point of observation and 
the contour $C$ of the Wilson loop, as an interaction of an antisymmetric 
tensor field, possessing a multivalued action, with the string world-sheet. 
In this respect, summation over surfaces bounded by $C$, i.e. summation over 
string world-sheets, could be treated as a summation over the 
branches of this multivalued action.

However, as it has already been discussed in Ref. [2], due to the Debye 
screening in the monopole gas, 
the photon in the 
model under study acquires a mass equal to   
the mass of the field $\chi$, following 
from the low-energy expansion of Eq. (2), $m=\frac{2\pi}{e}
\sqrt{2\zeta}$. This means that this mass should be taken into account 
in the monopole interaction from the very beginning, i.e. the Coulomb 
interaction in Eq. (1) should be replaced by the Yukawa one with the 
mass $m$, which leads to the following modification of Eq. (1) 

$${\cal Z}=\sum\limits_{N=1}^{+\infty}\sum\limits_{q_a=\pm 1}^{}
\frac{\zeta^N}{N!}
\prod\limits_{i=1}^N 
\int d{\bf z}_i\exp\left[-\frac{\pi}{2e^2}\sum\limits_{a\ne b}^{}
\frac{q_a q_b}{\left|{\bf z}_a-{\bf z}_b\right|}{\rm e}^{-m\left|
{\bf z}_a-{\bf z}_b\right|}
\right].$$
This expression could be again represented 
as a partition function of the sine-Gordon 
model of an auxiliary scalar field with the mass $m\sqrt{2}$, 

$${\cal Z}=\int {\cal D}\varphi\exp\left\{-\int d{\bf x}\left[\frac12 
(\partial_\mu\varphi)^2+\frac{m^2}{2}\varphi^2-2\zeta\cos
\left(\frac{2\pi}{e}
\varphi\right)\right]\right\}. \eqno (3)$$

In the next Section, we shall study the effects bringing about by 
the additional mass term in Eq. (3) both to the 3D  
compact QED and UCST. First, we shall 
cast Eq. (3) into the form of an integral 
over monopole densities, after which we shall study the influence of the 
mass term to the UCST and treat the obtained string theory both in 
the low- and high-energy limits. The latter one differs drastically 
from such a limit corresponding to the case when the photon is considered 
to be massless.

In Section 3, we shall find the mass operator of a  
string in the low-energy 
limit of the UCST, corresponding to the 
4D compact QED. 

We shall end up with a short Conclusion. 

\vspace{6mm}
{\large\bf 2. Accounting for the Finiteness of the Photon Mass in 
3D Compact QED and UCST}
\vspace{3mm}

Let us start with a representation of Eq. (3) as an integral 
over monopole densities. To this end, one should introduce into Eq. (3) a 
unity of the form 

$$\int {\cal D}\rho{\,} 
\delta\left(\rho\left({\bf x}\right)-\sum\limits_a^{} q_a\delta \left(
{\bf x}-{\bf z}_a\right)\right)=\int {\cal D}\rho{\,}
{\cal D}\mu{\,}\exp\left\{i\left[
\int d{\bf x}
\mu\rho-\sum\limits_a^{}q_a\mu\left({\bf z}_a\right)\right]\right\},$$
which, after the change of variables,  
$\mu\to\psi\equiv\frac{2\pi}{e}\varphi-\mu$,   
and integrations over the fields $\varphi$ and $\psi$,  
yields the following representation for the partition function 

$${\cal Z}=\int {\cal D}\rho\exp\left\{-\left[\frac{\pi}{2e^2}
\int d{\bf x} d{\bf y} \rho\left({\bf x}\right)\frac{{\rm e}^{-m\left|
{\bf x}-{\bf y}\right|}}{\left|
{\bf x}-{\bf y}\right|}\rho\left({\bf y}\right)+V\left[\rho\right]\right]
\right\},$$
where

$$V\left[\rho\right]\equiv\int d{\bf x}\left\{\rho\ln\left[\sqrt{1+
\left(\frac{\rho}{2\zeta}\right)^2}+\frac{\rho}{2\zeta}\right]
-2\zeta\sqrt{1+
\left(\frac{\rho}{2\zeta}\right)^2}\right\}$$
is a parabolic-type effective potential, whose asymptotic behaviours 
at $\rho\ll\zeta$ and $\rho\gg\zeta$ read

$$V\left[\rho\right]\longrightarrow\int d{\bf x}\left( 
-2\zeta+\frac{\rho^2}{4\zeta}\right)$$
and 

$$V\left[\rho\right]\longrightarrow\int d{\bf x}\left[ 
\rho\left(\ln\frac{\rho}{\zeta}-
1\right)\right]$$
respectively. Thus, one can represent 
3D compact QED as a nonlocal theory of the monopole densities 
with the Yukawa interaction and a certain effective potential. 

Let us now proceed to the UCST, i.e. to the dual formulation of 
the Wilson loop

$$\left<W(C)\right>=\left<\exp\left[i\sum\limits_a^{}q_a\eta\left(
{\bf z}_a\right)\right]\right>=$$

$$=\int {\cal D}\varphi\exp\left\{
-\int d{\bf x}\left[\frac12(\partial_\mu \varphi)^2+\frac{m^2}{2}
\varphi^2-
2\zeta\cos\left(\frac{2\pi}{e}\varphi+\eta\right)\right]\right\},$$
where $\eta\left({\bf x}\right)\equiv\frac12\int d\sigma_\mu\left({\bf y}
\right)\frac{\left({\bf x}-{\bf y}\right)_\mu}{\left|{\bf x}-{\bf y}
\right|^3}$ 
stands for the solid angle formed by the point ${\bf x}$ and the 
contour $C$. This could be done similarly to Ref. [3] 
by introduction of an auxiliary integration 
over an antisymmetric tensor field as follows 

$$\left<W(C)\right>=\int {\cal D}\phi {\cal D}B_{\mu\nu}\exp\Biggl\{
-\int d{\bf x}\Biggl[\frac{9}{e^2}B_{\mu\nu}^2-\frac{3i}{2\pi}\phi
\varepsilon_{\mu\nu\lambda}\partial_\mu B_{\nu\lambda}-$$

$$-2\zeta\cos\phi+
\zeta\left(\phi-\eta\right)^2\Biggr]+
i\int d\sigma_{\mu\nu}B_{\mu\nu}\Biggr\},$$
where 
$\phi\equiv\frac{2\pi}{e}\varphi+\eta$, $d\sigma_{\mu\nu}=
\varepsilon_{\mu\nu\lambda}d\sigma_\lambda$, 
and further elimination of the field $\phi$, which has no more 
kinetic term in the action.    
However now, contrary to Ref. [3], due to the presence of the 
mass term $\zeta\phi^2$ of the field $\phi$ in the Lagrangian,  
the saddle-point equation for this field has no more simple solutions 
and could be solved only in the low- or in the high-energy limits.

In the first case, i.e. when $\phi\ll 1$, the Wilson loop takes 
the following form 

$$\left<W(C)\right>=\exp\left[-\frac{\pi\zeta}{12}\int 
d\sigma_{\mu\nu}\left({\bf x}\right)\int d\sigma_{\mu\nu}\left(
{\bf y}\right)\frac{1}{\left|{\bf x}-{\bf y}\right|}\right]\cdot$$

$$\cdot\int {\cal D}B_{\mu\nu}\exp\left[-\int d{\bf x}\left(
\frac{3}{16\pi^2\zeta}H_{\mu\nu\lambda}^2+
\frac{9}{e^2}B_{\mu\nu}^2\right)+\frac{i}{2}\int d\sigma_{\mu\nu}
B_{\mu\nu}\right], \eqno (4)$$
where $H_{\mu\nu\lambda}=\partial_\mu B_{\nu\lambda}+\partial_\lambda 
B_{\mu\nu}+\partial_\nu B_{\lambda\mu}$ is the strength tensor of the 
field $B_{\mu\nu}$. One can see that except for the 
$B_{\mu\nu}$-independent part of the action, 

$$\frac{\pi\zeta}{12}\int 
d\sigma_{\mu\nu}\left({\bf x}\right)\int d\sigma_{\mu\nu}\left(
{\bf y}\right)\frac{1}{\left|{\bf x}-{\bf y}\right|}, \eqno (5)$$
Eq. (4) is just the 
low-energy limit of the UCST, obtained in Ref. [3] without 
accounting for the finiteness of the photon mass. 

It is easy to 
calculate the string tension of the Nambu-Goto term  
and the inverse bare coupling constant 
of the rigidity term bringing about by Eq. (5). 
To this end, we shall make use of the results of Ref. [4] 
and introduce a dimensionless cutoff $L\equiv
\frac{e^3}{\sqrt{\zeta}}\sim
\frac{e}{\varphi^{{\rm extr.}}}$, which is much larger than unity 
in the low-energy limit of the field $\varphi$ under consideration. 
Then the 
contributions of Eq. (5) to the quantities mentioned 
above read as follows 

$$\Delta\sigma=\frac{\pi^2\zeta^{\frac23}}{3}L\sim\frac{\pi^2
\zeta^{\frac16}e^3}{3} \eqno (6)$$
and 

$$\Delta\frac{1}{\alpha_0}=-\frac{\pi^2}{144}L^3\sim-\frac{\pi^2}{144}
\frac{e^9}{\zeta^{\frac32}} $$
respectively. 
 
Another case, when it is also 
possible to  
solve the saddle-point equation, is the case $\phi\gg 1$. This could be 
done by making use of the iterative procedure, 
which in the first order yields the following expression for the 
Wilson loop 

$$\left<W(C)\right>=\int {\cal D}B_{\mu\nu}\exp\left\{-\int d{\bf x}
\left[\frac{3H_{\mu\nu\lambda}^2}{8\pi^2\zeta}+\frac{9}{e^2}B_{\mu\nu}^2-
2\zeta\cos\left(\frac{i}{4\pi\zeta}\varepsilon_{\mu\nu\lambda}
H_{\mu\nu\lambda}+\eta\right)\right]\right\}. \eqno (7)$$
One can see that Eq. (7) 
is quite different from the expression for the UCST partition function
corresponding to the case when the photon is 
considered to be massless, studied in Ref. [3]. 

\vspace{6mm}

{\large\bf 3. Mass Operator 
of the Confining String in the Low-Energy Limit}
\vspace{3mm}

In the low-energy limit, the kinetic 
term of the field $B_{\mu\nu}$ in Eq. (4) vanishes, and we are left 
with the string described by the action (5), interacting with a constant 
antisymmetric tensor field with a certain Gaussian measure, which should 
be eventually averaged over. For simplicity, one can take into account 
only the Nambu-Goto term in the expansion of the action (5), whose 
string tension is given by Eq. (6).   

The Nambu-Goto string interacting with a constant antisymmetric tensor field 
with a Gaussian measure appears also in the low-energy limit of the 
4D UCST, studied in Ref. [5], where the finiteness of the photon mass 
has not been taken into account. In this case, it appears however not  
in the simplest approximation, when one treats the field $B_{\mu\nu}$ 
as a constant one. Namely, let us consider the low-energy expression 
for the partition function of the 4D UCST (i.e. the dual expression for 
the Wilson loop in the confining phase of 4D compact QED),  

$$\left<W(C)\right>=\int {\cal D}B_{\mu\nu}\exp\left[-\int 
d^4x\left(\frac{1}{12\Lambda^2}H_{\mu\nu\lambda}^2+\frac{1}{4e^2}
B_{\mu\nu}^2\right)+i\int d\sigma_{\mu\nu}B_{\mu\nu}\right],$$
where $\Lambda$ stands for the UV momentum cutoff, and split the 
total field $B_{\mu\nu}$ into an $x$-independent 
background part, $b_{\mu\nu}$, 
and a quantum fluctuation, $h_{\mu\nu}(x)$. Since due to the Hodge 
decomposition theorem [6], $h_{\mu\nu}(x)$ could be always 
represented in the form 
$h_{\mu\nu}=\partial_\mu A_\nu-\partial_\nu A_\mu+\partial_\lambda 
C_{\lambda\mu\nu}$,  
where $A_\mu$ and $C_{\lambda\mu\nu}$ stand for some vector 
and an antisymmetric rank-3 tensor respectively, the term 
$\frac{1}{2e^2}\int d^4x b_{\mu\nu}h_{\mu\nu}$ during this splitting  
vanishes by virtue of partial integration, and we arrive at the 
following expression for the Wilson loop

$$\left<W(C)\right>=
\int\limits_{-\infty}^{+\infty}\prod
\limits_{\mu,\nu=1;\mu\ne\nu}^4 
db_{\mu\nu}\exp\left(-\frac{V^{(4)}b_{\mu\nu}^2}
{4e^2}+i\int d\sigma_{\mu\nu}b_{\mu\nu}\right)\cdot$$

$$\cdot\exp\left[-\int 
d\sigma_{\alpha\beta}(x)\int d\sigma_{\lambda\rho}(x')D_{\alpha\beta, 
\lambda\rho}
(x-x')\right], $$
where $V^{(4)}$ is the four-volume of observation, and 
$D_{\alpha\beta, \lambda\rho}(x-x')$ stands for the propagator of the 
massive Kalb-Ramond field $h_{\mu\nu}$. What is important for us here is 
not the explicit form of this propagator, but the fact that  
the leading term of the derivative expansion 
of the action $\int 
d\sigma_{\alpha\beta}(x)\int d\sigma_{\lambda\rho}(x')D_{\alpha\beta, 
\lambda\rho}
(x-x')$ is the Nambu-Goto one with the string tension [5]  
$\sigma=\frac{\Lambda^2}{4\pi}K_0\left(\frac{1}{4e\exp\left(
\frac{{\rm const.}}{e^2}\right)}\right)$,  
where $K_0$ stands for the Macdonald function. Thus we have again 
arrived at a theory of the Nambu-Goto string interacting with the 
constant antisymmetric 
tensor field $b_{\mu\nu}$, which possesses the quadratic action 
$S\left[b_{\mu\nu}\right]=\frac{V^{(4)}b_{\mu\nu}^2}{4e^2}$. 

In what follows, 
for concreteness, we shall study this very theory, rather than the 
analogous one, 
described in the first paragraph of the present Section, which follows 
from 3D compact QED, and  
find the mass operator of the confining string in this theory. To this end, 
we shall make use of the result of Ref. [7], where the mass operator 
of the Nambu-Goto string in the external constant electromagnetic field 
has been found. The difference of our case from the one studied 
in Ref. [7] is the 
necessity of performing the average over the field $b_{\mu\nu}$. In 
this way, we find the following expression for the operator of the 
square of mass 
of the confining string 

$$M^2=2\pi\sigma\left[\sum\limits_{n=1}^{+\infty}n:{\hat a}_n^\dag 
{\hat a}_n:-
\alpha(0)\right]\frac{\int\limits_{-\infty}^{+\infty}\prod
\limits_{\mu,\nu=1;\mu\ne\nu}^{4}db_{\mu\nu}\frac{{\rm e}^{-S\left[b_{\mu
\nu}\right]}}{1+\frac{2}{\sigma^2}\left(b_{12}^2+b_{13}^2+b_{14}^2+
b_{21}^2+b_{31}^2+b_{41}^2\right)}}{\int\limits_{-\infty}^{+\infty}\prod
\limits_{\mu,\nu=1;\mu\ne\nu}^{4}db_{\mu\nu}{\rm e}^{-S\left[b_{\mu
\nu}\right]}}, \eqno (8)$$
where the eigenvalues of the operator $\sum\limits_{n=1}^{+\infty}n:
{\hat a}_n^\dag 
{\hat a}_n:$ are equal to $1,2,...$, and $\alpha(0)<0$. In order to 
carry out the integral standing in the numerator on the R.H.S. of Eq. (8), 
one should split the infinite interval of integration over 
the absolute value of the six-dimensional vector into two pieces, from 
$0$ to $\frac{\sigma}{2e}\sqrt{\frac{V^{(4)}}{2}}$ and from 
$\frac{\sigma}{2e}\sqrt{\frac{V^{(4)}}{2}}$ to $+\infty$, and neglect 
in these two intervals the terms $\frac{2}{\sigma^2}
\left(b_{12}^2+b_{13}^2+b_{14}^2+
b_{21}^2+b_{31}^2+b_{41}^2\right)$ and $1$ in the integrand, 
respectively. Then Eq. (8) yields the following value of the operator 
of the square of mass of the confining string in the low-energy limit 

$$M^2=2\pi\sigma\left[\sum\limits_{n=1}^{+\infty}n:{\hat a}_n^\dag 
{\hat a}_n:-
\alpha(0)\right]\left\{1-\frac12\left[\left(\frac{\sigma^2V^{(4)}}{8e^2}
\right)^2+\frac{\sigma^2V^{(4)}}{8e^2}+
1\right]\exp\left(-\frac{\sigma^2V^{(4)}}{8e^2}
\right)\right\}, \eqno (9)$$
which could be checked to be always positive. In Eq. (9),  
the effect of the finiteness of the volume of observation has been taken 
into account. The value of $M^2$ in the limit $V^{(4)}
\to +\infty$ is obvious. 

\vspace{6mm}
{\large\bf 4. Conclusion}

\vspace{3mm}
In the present Letter, we have addressed two problems. In Section 2, 
we have studied the influence of the effects 
bringing about by the finiteness of the photon mass in the 3D compact 
QED to the representation of the partition function of this theory 
as an integral over monopole densities. As a result, we have obtained the 
nonlocal theory of these densities with the Yukawa interaction and 
a certain parabolic-type effective potential. Next, we have investigated 
the influence of the finiteness of the photon mass to the dual 
formulation of the Wilson loop, i.e. to the theory of confining strings. 
In the low-energy limit, the obtained string theory coincides 
(apart from the additional contribution (5) to the action) with the 
one corresponding to the case when the photon mass is not taken into 
account, whereas in the high-energy case, described by Eq. (7), 
these two theories are quite different from each other. 

In Section 3, we have found the mass operator of the confining string 
corresponding to the low-energy limit of the 4D compact QED, where 
the photon has been considered to be massless. In the final result (9), 
the dependence on the four-volume of 
observation has been explicitly presented.

\vspace{6mm}
{\large \bf 5. Acknowledgments}

\vspace{3mm}
The author is indebted to Prof. H. Kleinert for bringing his attention to 
some Refs. of [1] and 
Dr. H. Dorn and V.I. Shevchenko 
for bringing his attention to Ref. [6].  
He is also grateful to the theory group of the Quantum 
Field Theory Department of the Institut f\"ur Physik of the 
Humboldt-Universit\"at of Berlin for kind hospitality. 
This work is  
supported 
by Graduiertenkolleg {\it Elementarteilchenphysik}, Russian 
Foundation for Basic Research, project 96-02-19184, DFG-RFFI,
grant 436 RUS 113/309/0, and by the INTAS, grant No.94-2851.

\newpage
{\large\bf References}

\vspace{3mm}
\noindent
$[1]$~K.G. Wilson, Phys. Rev. D 10 (1974) 2445; T. Banks, R. Myerson, 
and J. Kogut, Nucl. Phys. B 129 (1977) 493; M. Peskin, Ann. Phys. 
113 (1978) 122; J.L. Cardy, Nucl. Phys. B 170 (1980) 369; 
T.A. De Grand and D. Toussaint, Phys. Rev. D 22 (1980) 2478; 
H. Kleinert, Gauge Fields in Condensed Matter, Vol. 1: Superflow 
and Vortex Lines. Disorder Fields, Phase Transitions (World Scientific 
Publishing Co., Singapore, 1989);  
P. Cea and L. Cosmai, Phys. Lett. B 249 (1990) 114;   
H. Kleinert, Phys. Lett. B 246 (1990) 127, Int. J. Mod. 
Phys. A 7 (1992) 4693, Phys. Lett. B 293 (1992) 168; 
K. Lee, Phys. Rev. D 48 (1993) 2493; E.T. Akhmedov, M.N. Chernodub, 
M.I. Polikarpov, and M.A. Zubkov, Phys. Rev. D 53 (1996) 2087; 
E.T. Akhmedov, JETP Lett. 64 (1996) 82.\\  
$[2]$~A.M. Polyakov, Nucl. Phys. B 120 (1977) 429.\\
$[3]$~A.M. Polyakov, Nucl. Phys. B 486 (1997) 23.\\
$[4]$~D.V. Antonov, D. Ebert, and Yu.A. Simonov, Mod. Phys. Lett. 
A 11 (1996) 1905.\\
$[5]$~M.C. Diamantini, F. Quevedo, and C.A. Trugenberger, Phys. Lett. 
B 396 (1997) 115; M.C. Diamantini and C.A. Trugenberger, preprints 
UGVA-DPT 1997/11-992 and hep-th/9712008 (1997).\\
$[6]$~See e.g. M. Nakahara, Geometry, Topology, and Physics 
(England, London, 1990).\\ 
$[7]$~B.M. Barbashov, V.V. Nesterenko, and A.M. Chervyakov, Teor. Mat. 
Fiz. 32 (1977) 336.

\end{document}